\documentstyle[aps,prb,preprint,epsf]{revtex}
\input psfig

\begin{document}
\setcounter{page}{1}
\date{\today}
\title{Spin gap behavior and charge ordering in $\alpha^\prime$-
NaV$_{2}$O$_{5}$ probed by light scattering}
\author{M. Fischer, P. Lemmens, G. Els, G. G\"untherodt}
\address{2. Physikalisches Institut, RWTH Aachen, Templergraben 55, 52056 Aachen, Germany}
\author{E. Ya. Sherman}
\address{Moscow Institute of Physics and Technology, 141700
Dolgoprudny, Moscow region, Russia}
\author{E. Morre, C. Geibel, F. Steglich}
\address{Max-Planck-Institut f\"ur Chemische Physik fester Stoffe, 01187 Dresden, Germany}

\maketitle
\begin{abstract}

We present a detailed analysis of light scattering experiments
performed on the quarter-filled spin ladder compound
$\alpha^\prime$-NaV$_{2}$O$_{5}$ for the temperature range
5~K$\le$T$\le$300~K. This system undergoes a phase transition into
a singlet ground state at T=34~K accompanied by the formation of a
super structure. For T$\leq$34~K several new modes were detected.
Three of these modes are identified as magnetic bound states.
Experimental evidence for charge ordering on the V sites is
detected as an anomalous shift and splitting of a V-O vibration at
422~cm$^{-1}$ for temperatures above 34~K. The smooth and
crossover-like onset of this ordering at T$_{\rm CO}$= 80~K is
accompanied by pretransitional fluctuations both in magnetic and
phononic Raman scattering. It resembles the effect of stripe order
on the super structure intensities in La$_2$NiO$_{4+\delta}$.

\end{abstract}
\pacs{PACS: Spin-Peierls transition,
$\alpha^\prime$-NaV$_{2}$O$_{5}$, Raman scattering, charge
ordering}

\section{INTRODUCTION}
Low dimensional quantum spin systems have recently attracted
strong experimental as well as theoretical interests. Despite
their principally simple crystal structure they exhibit a variety
of unusual ground states and excitations. One of these unusual
phenomena which are very specific to one-dimensional spin systems
is the spin-Peierls transition. The nature of this transition is
an instability of a spin-1/2 Heisenberg chain against a dimerized
ground state due to a coupling of the spins to the lattice. This
instability appears as a well defined thermodynamic phase
transition at a finite temperature. Thereby the lattice undergoes
a distortion leading to a small structural change and thus to a
super structure. On the other hand, two neighboring spins dimerize
leading to a singlet ground state and a singlet-triplet gap in the
magnetic excitation spectrum \cite{pytte,cross,bray}.

If an additional frustration due to next nearest neighbor exchange
interaction is present, as, e.g., in the inorganic spin-Peierls
compound CuGeO$_{3}$ \cite{hase,nishi,emery,boucher}, a magnetic
singlet bound state exists that is combined of two triplet
excitations. This bound state is indeed observed in Raman
experiments due to exchange light scattering (no spin flip, i.e.
$\Delta$S=0) \cite{kuroe,mutu,lemmens}. In spin ladders multiple
magnetic bound states of triplet or singlet character are expected
\cite{kotov98}. Here, with increasing spin frustration the binding
energy and the number of bound states increases. It is proposed
that the quantum phase transition from a gapless into a quantum
disordered state with gapped elementary excitations is
characterized by the condensation of many particle bound states
\cite{kotov98}.

Recently, it has been proposed that
$\alpha^\prime$-NaV$_{2}$O$_{5}$ might be the second inorganic
spin-Peierls compound with a transition temperature T$_{\rm
SP}$=34~K \cite{isobe}. Indeed, susceptibility data of
$\alpha^\prime$-NaV$_{2}$O$_{5}$ can be fit very well by a
one-dimensional Heisenberg chain model yielding an exchange
interaction of J=480 and 560~K determined for temperatures below
and above T$_{\rm SP}$, respectively \cite{isobe,weiden}. For
T$\le$T$_{\rm SP}$ an isotropic drop in the susceptibility
corresponding to the opening of a singlet-triplet gap of
$\Delta_{\rm ST}$=85~K has been observed. X-ray measurements
\cite{fuji} as well as Raman scattering \cite{weiden} proved the
existence of a super structure formation.

However, recent X-ray structure data analysis at room temperature
is not in favor of the previously reported non-centrosymmetric
structure \cite{galy} and of the picture of a one-dimensional
Heisenberg chain. They clearly give evidence for the
centrosymmetric space group (D$^{13}_{2h}$) leading to only one
type of V site with a formal valence +4.5 in this compound
\cite{roth,smolinski}. The V sites then form a quarter-filled
ladder running along the b-axis with rungs along the
crystallographic a-axis. For low energy scales the quarter-filled
ladder can effectively be mapped onto a s=1/2-chain along the
b-axis as there is a considerable large charge transfer gap. Both
the estimated charge transfer gap and the exchange coupling J
agree well with the experimental observations
\cite{smolinski,horsch}. In principle, it is possible that this
effective chain shows all the features one may expect from an
ordinary one-dimensional spin chain, including a spin-Peierls
instability.

On the other hand, for quarter-filled bands there are several
other instabilities such as spin density wave or charge density
wave transitions. As shown by Seo and Fukuyama \cite{seo97}
intersite Coulomb interaction is the most important parameter in
this sense. It introduces charge disproportionation or other
instabilities of the system. Therefore new scenarios for the phase
transition in $\alpha^\prime$-NaV$_{2}$O$_{5}$ are discussed. One
possibility is to assume a primary charge order of the V sites
into V$^{4+}$ with s=1/2 and V$^{5+}$ with no spin on either leg
of the ladder leading to linear spin chains in b-direction. This
charge ordering is followed by a secondary spin-Peierls transition
\cite{thalmeier}.

A different approach has been chosen by Mostovoy and Khomskii
\cite{mostovoy}. Starting with an electronic Hamiltonian including
not only hopping terms and on-site Coulomb repulsion but also
inter-site Coulomb interaction a charge ordering of the
V$^{4+}$/V$^{5+}$ in a zig-zag structure with diagonal dimers is
obtained. This also results in an alternation of the spin exchange
constant along the b-direction and thus to a spin gap. The charge
ordering is thereby driven by the interaction with the lattice
distortions and intersite Coulomb interaction. A similar ansatz
used by Seo and Fukuyama \cite{seo} neglecting lattice distortions
also leads to a dimer formation along the diagonal of the ladder
as the energetically preferred ground state of
$\alpha^\prime$-NaV$_{2}$O$_{5}$. From the experimental point of
view diagonal charge ordering is compatible with a splitting of a
triplet branch observed in neutron scattering
\cite{yoshihama,gros} and Raman scattering results presented here.

\section{EXPERIMENTAL}
Raman experiments were performed using an Ar$^{+}$-ion laser
($\lambda$=514~nm) and a HeNe laser ($\lambda$=632.8~nm) as
excitation source and a Dilor XY spectrometer. A nitrogen cooled
CCD detector was used for detection of the scattered light. The
single crystals were grown using a self flux method. Two different
single crystals were available: One that showed a clear drop at
T$_{\rm SP}$=34~K in susceptibility measurements and another one
with a slightly broadened transition due to a small Na deficiency.
These samples will be called here sample 1 and sample 2,
respectively. The crystals are thin platelets of several mm
diameter in the (ab) plane. Therefore all scattering geometries
with polarizations within the (ab) plane were easy accessible by
light scattering experiments. Other geometries could be probed
using micro-Raman at room temperature. The samples were cooled in
exchange gas of a standard He-bath cryostat. With this setup light
scattering in a temperature range 5~K$\le$T$\le$~300~K was
investigated.

\section{RESULTS AND DISCUSSION}
For the understanding of the observed phase transition and further
possible instabilities in $\alpha^\prime$-NaV$_{2}$O$_{5}$ it is
important to clarify the high temperature phase. As mentioned
above X-ray determination of the crystal structure gives doubtless
evidence for only one type of V site at room temperature. This is
also confirmed in polarized Raman scattering from phonons. Factor
group analysis as described, e.g., in Ref. \onlinecite{rousseau}
gives the number of symmetry allowed normal modes in Raman and
infrared measurements. There are two formula units of
$\alpha^\prime$-NaV$_{2}$O$_{5}$ per unit cell. Therefore one
expects 45 modes in Raman scattering. For the non-centrosymmetric
space group (C$^{7}_{2v}$) one obtains $\Gamma=$15 A$_1$+8
A$_2$+7B$_1$+15 B$_2$. The A$_1$ modes should be observable in all
"parallel light polarizations" in Raman spectroscopy as well as
for {\bf E}$\parallel$a in infrared measurements. Indeed, as for
for this space group there is no center of inversion symmetry,
there is no distinction between even and odd modes. The B$_{1}$
modes are observable in (ab) polarization as well as for {\bf
E}$\parallel$b. The A$_{2}$ and the B$_{2}$ modes with selection
rules (bc), and (ac), {\bf E}$\parallel$c, respectively, are not
observable in our experiment because these polarizations are not
accessible with our crystals.

\begin{figure}
\centerline{\psfig{file=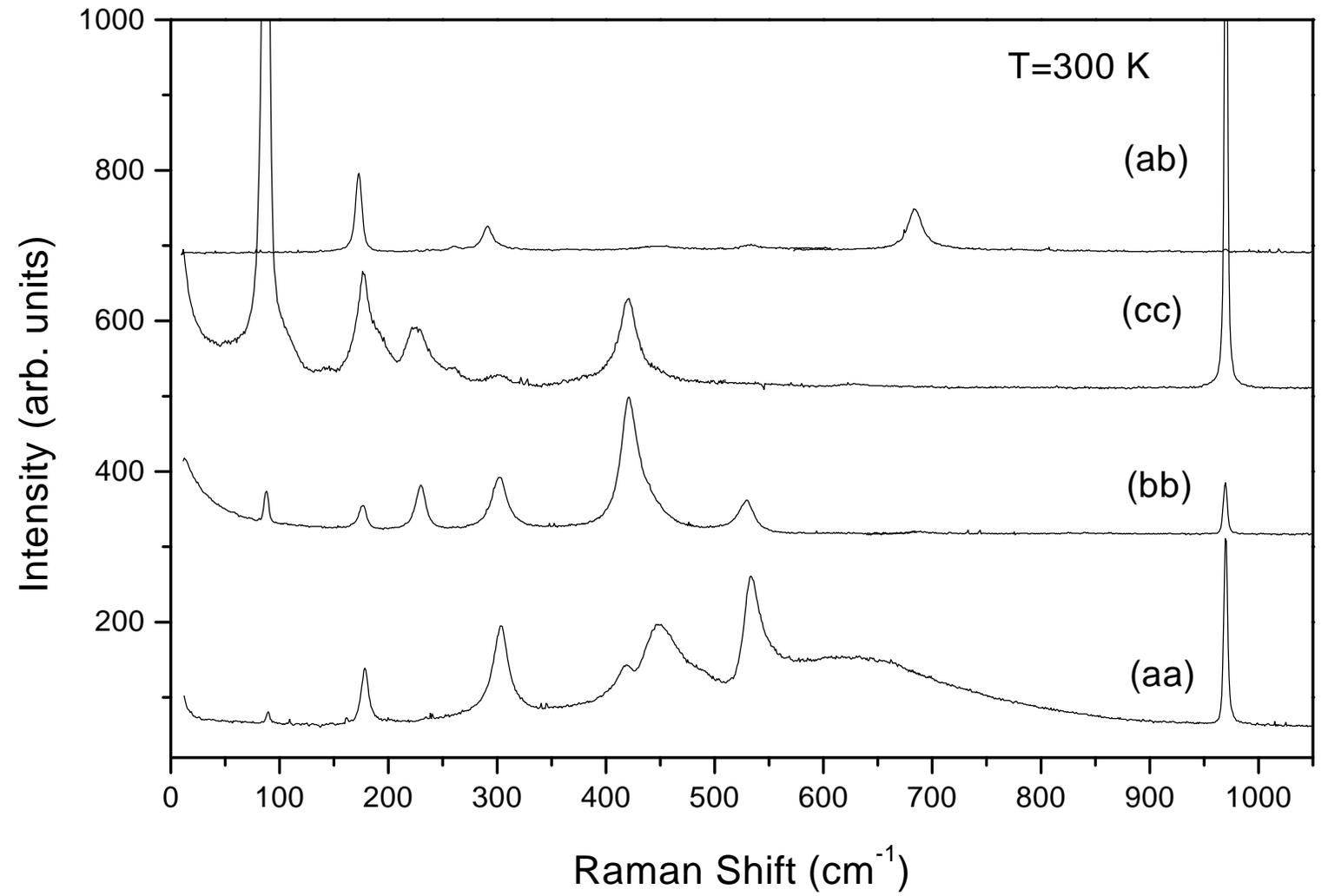,height=16cm,rheight=16cm}}
\caption{Room temperature Raman spectra of a
$\alpha^\prime$-NaV$_{2}$O$_{5}$ single crystal. Measurements in
(cc) polarization were carried out using a micro-Raman setup.}
\label{f1}
\end{figure}

For the centrosymmetric space group (D$^{13}_{2h}$) there is a
center of inversion symmetry leading to a clear distinction
between Raman and infrared active modes. One obtains 24 Raman
active phonon modes: $\Gamma$= 8 A$_{g}$+ 3 B$_{1g}$ + 8 B$_{2g}$
+5 B$_{3g}$. The residual 18 modes are infrared active. The
A$_{g}$ modes are observable in parallel light polarizations, i.e.
(aa), (bb), and (cc). The B$_g$ modes, on the other hand, appear
in "crossed polarizations". In our experiment the (ab) scattering
configuration with 3 B$_{1g}$ modes expected is accessible. Figure
\ref{f1} shows typical room temperature spectra for the
geometries. Table 1 summarizes the modes and compares them to the
predictions for the different space groups. It can be stated that
the centrosymmetric space group is consistent with the Raman
results. Nevertheless, it can not be excluded that due to matrix
elements effects, i.e. weak polarizability of certain modes, we
were not able to detect all of them. However, infrared
measurements as shown in Ref. \onlinecite{damanavo} are also in
agreement with the above interpretation of the data, especially as
there is no intermixing between Raman and infrared modes. This
symmetry analysis is valid for temperatures above T$_{\rm CO}$=
80~K. Changes observed at lower temperatures will be discussed
below. Our results are in contrast to an investigation of
Golubschik et al. favoring the non-centrosymmetric space group
(C$^{7}_{2v}$) \cite{golub}.

Considering the room temperature spectra one finds some unusual
features. The first striking point is the occurrence of
quasielastic scattered light in (bb) polarization, i.e. with the
polarization of the incoming and scattered light along the
effective chain direction. This scattering is not observed in (aa)
spectra. It vanishes upon cooling the sample below T=100~K. It
cannot be explained by the Bose-factor. Indeed, in CuGeO$_{3}$ a
similar quasielastic scattering contribution was found with
polarization parallel to the dominant exchange path (chain
direction) \cite{yamada,kuroea} and has been assigned to
fluctuations of the energy density of the spin system, a feature
which is very general for 1D quantum spin systems . Furthermore,
the phonons both in (aa) and (bb) geometry are broadened and show
an asymmetric line shape in the frequency range from $\sim$~250 to
700~cm$^{-1}$. This frequency regime corresponds roughly to
energies of one to two times J ($\approx$ 320-400 cm$^{-1}$)
pointing to a non-negligible spin-phonon interaction in this
system.


\subsection{Continuum scattering above T$_{\rm SP}$}
\begin{figure}
\centerline{\psfig{file=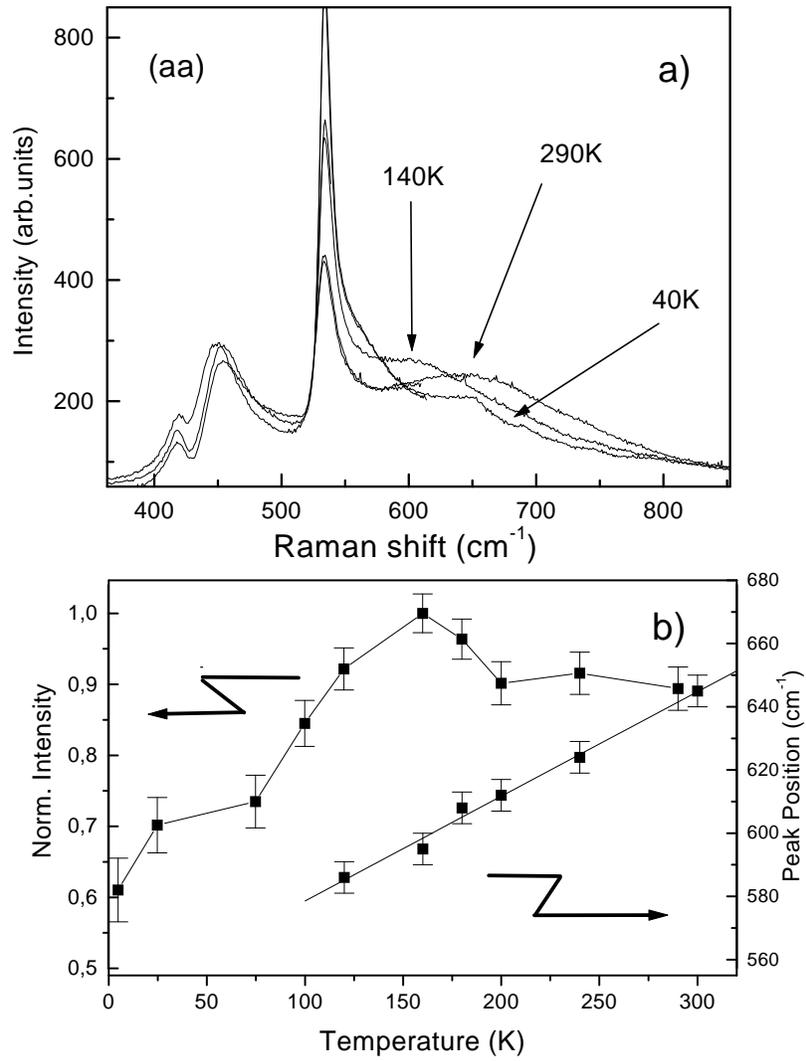,height=16cm,rheight=16cm}}
\caption{a) Broad scattering contribution in (aa) polarization of
$\alpha^{\prime}$-NaV$_{2}$O$_{5}$ at 290, 140, and 40~K. b)
Normalized integrated intensity of the scattering contribution
between 565 and 950 cm$^{-1}$ and peak position of the maximum as
a function of temperature.} \label{f3}
\end{figure}

The most remarkable signal in the room temperature spectra is
nevertheless the broad Gaussian-like band of scattering ranging
from 250 to 900~cm$^{-1}$ and centered at about 650~cm$^{-1}$.
Fig. \ref{f3}a shows typical spectra in (aa) scattering geometry
for several temperatures. Upon cooling to lower temperatures two
effects are observed. First, the center of this band shifts toward
lower frequencies. The shift is displayed in Fig. 2b). Second, the
intensity of the band decreases in favor of the phonon intensity
at 533~cm$^{-1}$.

This shift of the broad band is nearly linear in temperature down
to 100~K. Below 100~K the intensity is too small to be analyzed as
the Gaussian-like shape has disappeared. Instead an asymmetric
tail of the 533~cm$^{-1}$ phonon toward higher energies remains.
This is a remarkable change in the shape of this broad excitation.
In addition, its intensity shows a peculiar temperature
dependence. A maximum at T=150K and a consequent drastic drop for
lower temperatures is observed as it is shown in Fig. \ref{f3}b).
The phonon at 533~cm$^{-1}$, on the other hand, shows a clear
increase in intensity with decreasing temperature. To summarize,
there is a redistribution of spectral weight for temperatures
below 150~K suppressing the continuum scattering in favor of the
phonon scattering.

We will discuss several scenarios to explain these experimental
findings. First, electric dipole transitions between split crystal
field levels, second a two-magnon scattering mechanism and third
possible electron-phonon coupled modes observed in
antiferromagnets.

Electric dipole transitions between crystal field split levels of
d-electron states as the origin of the broad scattering
contribution were suggested by Golubschik et al. \cite{golub}.
These transitions between different crystal field levels should be
discrete. The broad continuum-like shape observed at high
temperatures may due to temperature smearing. Upon decreasing the
temperature the discrete nature of the excitations should be
recovered. However, this is not observed. In addition, independent
of the lineshape, the intensity should not show the observed
drastic dependence on temperature.

Two-magnon excitations can also lead to broad scattering bands in
Raman scattering. The corresponding Raman process is a
frustration-induced four-spinon continuum as observed in
CuGeO$_{3}$ \cite{mutu,brenig}. However, the polarization
selection rules allow its existence only for polarizations of the
incident and scattered light parallel to the dominant exchange
path \cite{mutu}. In addition, the observed energy range of the
band (250-850~cm$^{-1}$) is not compatible with the range expected
for an exchange coupling constant J=480-560~K in
$\alpha\prime$-NaV$_{2}$O$_{5}$. With this J value a maximum
centered around 1200-1400~cm$^{-1}$ would be expected.

There are, on the other hand, well comparable experimental results
discussed as electron-phonon coupled modes in antiferromagnets
(for an overview see Ref. \onlinecite{lockw}). One of the model
systems in this respect is FeCl$_{2}$ with a N\'eel temperature
T$_{\rm N}$=23.5~K. In this compound an E$_g$-phonon at
150~cm$^{-1}$ shows a line shape asymmetry toward higher energy
and a broadening for T~$\ge$~T$_{\rm N}$. For T~$\le$~T$_{\rm N}$
the phonon sharpens in linewidth and a broad Gaussian-like
excitation band is observed on the high frequency side that shifts
towards higher frequencies upon cooling further down well below
T$_{\rm N}$. This band is due to a resonant electron-phonon
interaction between a Fe$^{2+}$ d-level and the E$_{g}$ phonon.
The shift of the maximum observed in
$\alpha\prime$-NaV$_{2}$O$_{5}$ is opposite to these observations
in FeCl$_{2}$. This difference in the two classes of materials
concerning the electron-phonon coupled modes is connected with the
strong spin fluctuations in low dimensions. In the dimerized state
of a low dimensional spin system the spin-spin correlation length
decays exponentially and is therefore always smaller than in the
homogeneous state with algebraically decaying correlations. This
decrease of the correlation length is just opposite to the
behavior of an antiferromagnet cooled below the N\'eel
temperature. This leads to the reversed temperature dependence of
the electron-phonon coupled modes in the two different materials.

More general, in systems with low energy excitations electronic
satellites might appear on the high energy side of a phonon if the
low energy excitations and the phonon couple to the same
electronic state. This has been observed in Raman scattering on
Tl$_{2}$Ba$_{2}$CuO$_{6}$ \cite{misochko}. As the broad band in
$\alpha^\prime$-NaV$_{2}$O$_{5}$ is observed only for (aa)
polarization it serves as a strong experimental evidence that
there are additional low energy excitations in the a-direction,
i.e. perpendicular to the ladder legs. Further theoretical
investigations of this phenomenon are desirable, especially
concerning its peculiar temperature dependence, as it will give
valuable insight in the microscopic couplings between electronic
and phononic degrees of freedom in the high temperature phase of
this compound.

\subsection{Magnetic Bound States}
\begin{figure}
\centerline{\psfig{file=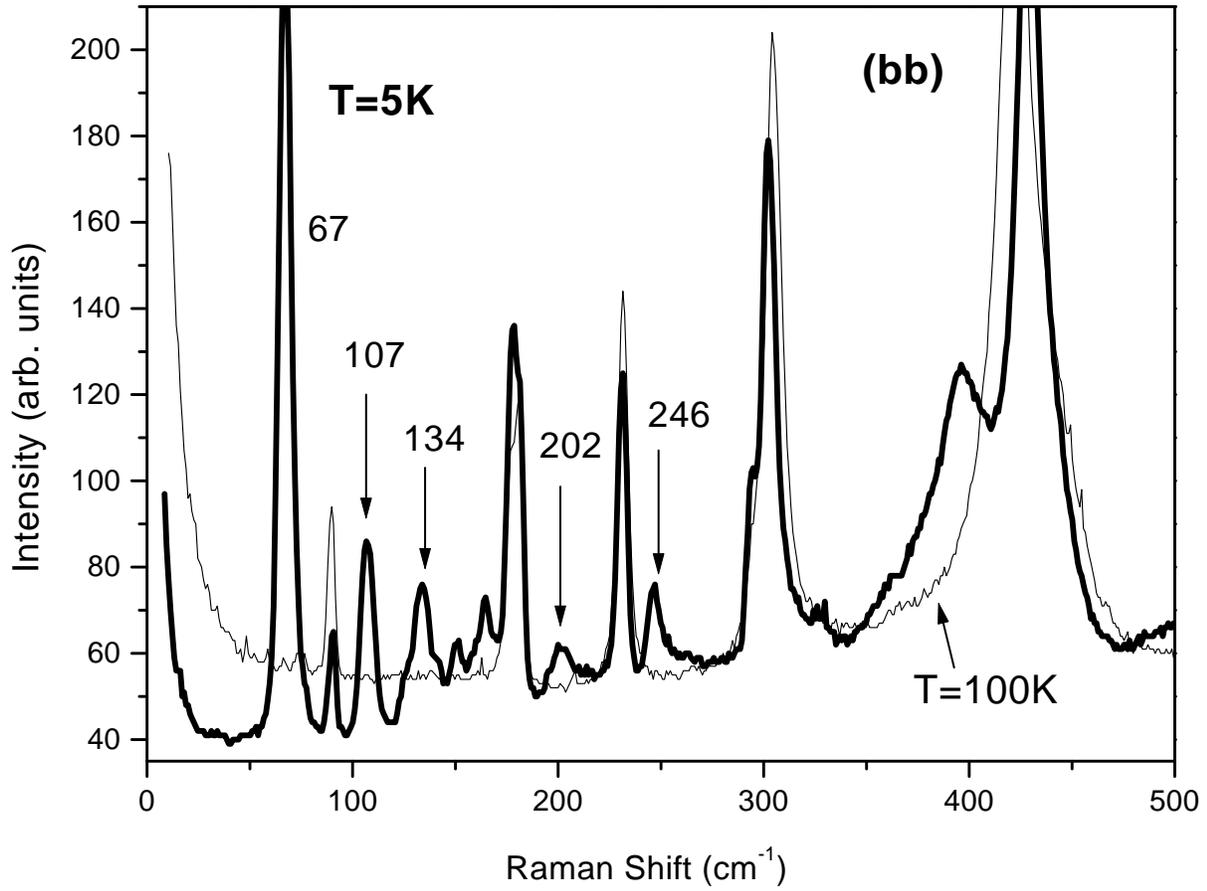,height=14cm,rheight=14cm}}
\caption{Low temperature modes of $\alpha^\prime$-NaV$_{2}$O$_{5}$
in (bb) scattering geometry:  transition-induced modes with
frequencies below 500~cm$^{-1}$. The thick (thin) line corresponds
to measurements at T=5K (100K). Note the drop in background
intensity for frequencies smaller than 120~cm$^{-1}$ corresponding
to 2$\Delta_{ST}$.} \label{f4a}
\end{figure}

\begin{figure}
\centerline{\psfig{file=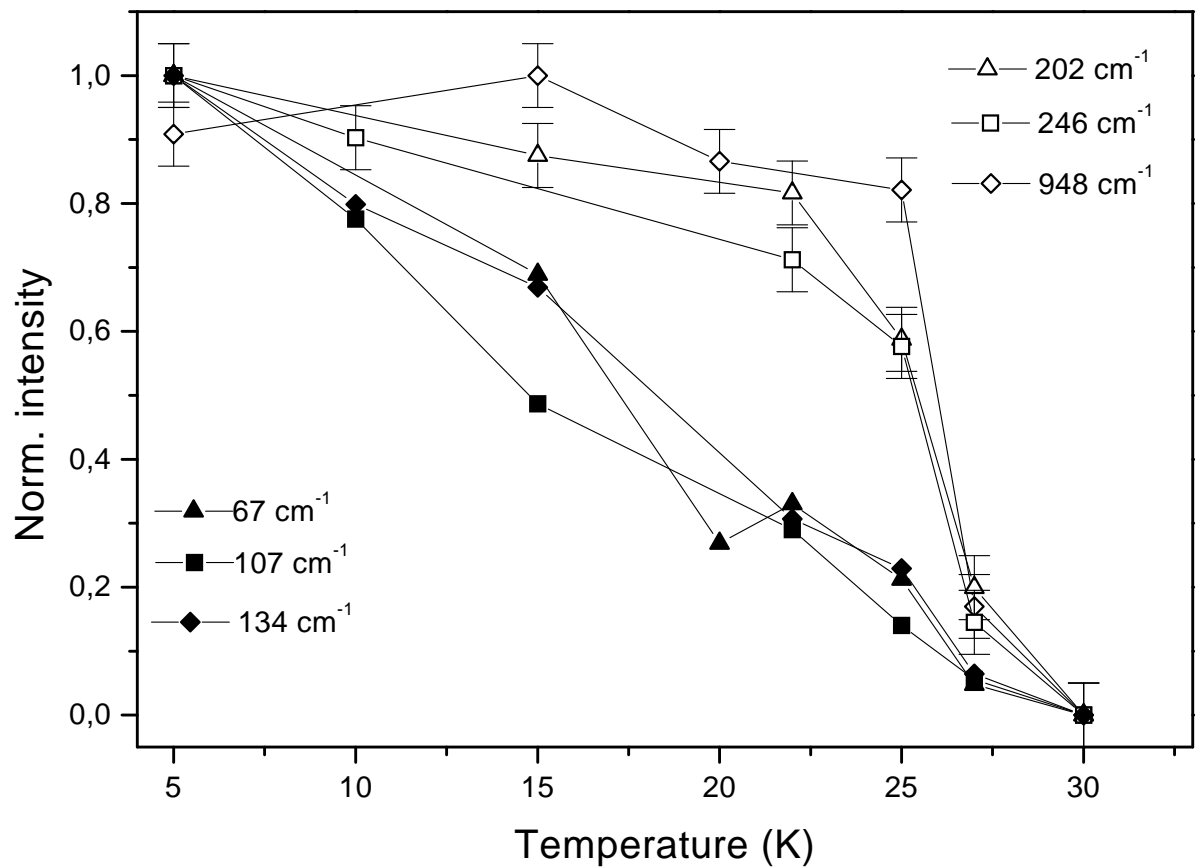,height=14cm,rheight=14cm}}
\caption{Temperature dependence of the normalized integrated
intensity of the transition-induced modes in
$\alpha^\prime$-NaV$_{2}$O$_{5}$. The three modes at 67, 107 and
134 cm$^{-1}$ (closed symbol) show a temperature dependency
different from the modes observed at higher frequencies.}
\label{f4b}
\end{figure}

In the low temperature phase (T$\leq$T$_{\rm SP}$) several new
modes do appear in polarizations along the "chains" (bb) (see Fig.
\ref{f4a}), and perpendicular to them (aa) as well as in crossed
polarization (ab) \cite{lemmens98}. A survey of the frequencies of
the additional modes is shown in Table 2. Comparing the
temperature dependence of the intensity, frequency and halfwidth
of the additional modes we come to a clear distinction between the
three modes at the lowest energies, i.e. at 67, 107, and 134
cm$^{-1}$ and the other transition-induced modes. The 67 cm$^{-1}$
and 134~cm$^{-1}$ modes appear in (aa), (bb) and (ab)
polarization, while the 107~cm$^{-1}$ mode is observed only in
(bb) polarization. Fig. \ref{f4b} shows the temperature dependence
of the normalized integrated intensity of the modes at 67, 107,
134, 202, 246, and 948 cm$^{-1}$ as observed in (bb) scattering
geometry. The latter three modes follow the behavior expected from
ordinary phonons that are folded from the zone boundary to the
zone center. Cooling down from high temperatures the intensities
rise sharply below T$_{\rm SP}$ and then saturate. The three low
frequency modes obviously do not behave like this. They increase
more gradually upon cooling with a smaller slope and show no
saturation in intensity toward the lowest temperature. Moreover,
these three modes get broader and soften toward lower frequencies
upon approaching T$_{\rm SP}$ from low temperatures
\cite{lemmens98}. So these modes really vanish above T$_{\rm SP}$
and are obviously closely connected to the opening of an energy
gap in the magnetic excitation spectrum. This is in contrast to
the modes at 202, 246, and 948 cm$^{-1}$ which do neither show any
broadening nor any shift in frequency upon approaching T$_{\rm
SP}$ from below. They can therefore  be assigned to phonons which
exist also above T$_{\rm SP}$, but become Raman active due to the
lowering of the symmetry for T$\le$T$_{\rm SP}$.

This conclusion is in close analogy to the situation in the
spin-Peierls compound CuGeO$_{3}$ where a similar distinction
between a mode at 30~cm$^{-1}$ and dimerization-induced folded
phonon modes could be made. The 30-cm$^{-1}$ mode was assigned to
a bound singlet state of two triplet excitations. The frustration
induced binding of the triplets leads to its energy below the
two-triplet continuum of states \cite{lemmens,bouz}. The linear
intensity increase with decreasing temperatures below T$_{\rm SP}$
results from the gradually developing order parameter as a
prere\-quisite of a composite state of two triplets. In this way
CuGeO$_{3}$ serves as a model system concerning the occurrence of
such magnetic bound states.

Nevertheless, the existence of such states is a quite general
feature of low dimensional spin systems with a gapped excitation
spectrum. We therefore assign the three modes at 67, 107, and 134
cm$^{-1}$ to magnetic singlet bound states. With $\Delta_{\rm
ST}$=85~K$\equiv$60 cm$^{-1}$ from susceptibility measurements
these three modes are situated below or near 2$\times$$\Delta_{\rm
ST}$. This gap value is also confirmed in our Raman measurements
as a drop in the background intensity below 120~cm$^{-1}$. The
number and selection rules of the singlet bound states, however,
differ from what has been observed in CuGeO$_{3}$ pointing to a
different ground state and excitation scheme in
$\alpha^\prime$-NaV$_{2}$O$_{5}$. We will discuss this point
below.

Now one may raise objections against the interpretation of the
modes as magnetic bound states. The first objection might be that
the 67-cm$^{-1}$ mode is comparable in frequency to the
singlet-triplet energy gap ($\Delta_{\rm ST}$=60~cm$^{-1}$) and
might therefore be just the one-magnon scattering which might be
allowed due to spin-orbit coupling. However, the g-value in
$\alpha^\prime$-NaV$_{2}$O$_{5}$ is even closer to 2 than in
CuGeO$_{3}$ \cite{isobe,schmidt98}, pointing to a negligible
orbital momentum in this compound. Furthermore, if the
67-cm$^{-1}$ mode would be interpreted as a one-magnon scattering,
the 134-cm$^{-1}$ mode may then be understood as the corresponding
two-magnon scattering \cite{kuroe98b}. This, however, would be
surprising since two-magnon scattering is strongly renormalized to
lower frequencies already in two dimensional spin systems
\cite{cottam}.

\begin{figure}
\centerline{\psfig{file=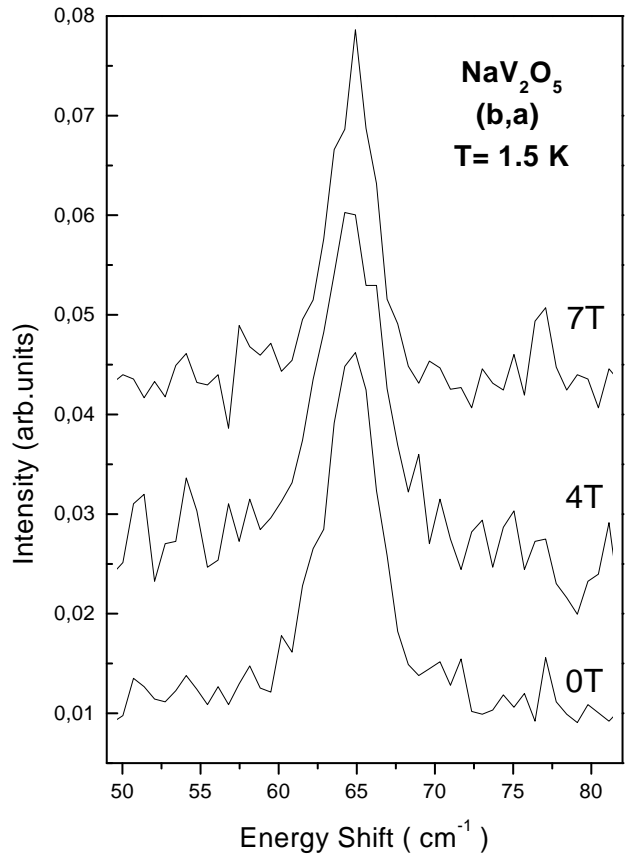,height=14cm,rheight=14cm}}
\caption{Behaviour of the 67-cm$^{-1}$ mode upon applying a
magnetic field of 0, 4 and 7~T. This measurement was performed
using a Brillouin Fabry-Perot spectrometer.} \label{f10}
\end{figure}

A very important test on relevant spin-orbit coupling is the
application of a magnetic field. This should lift the threefold
degeneracy of the triplet state, resulting in a splitting and a
shift of this mode. Fig.\ref{f10} shows the 67-cm$^{-1}$ mode
measured in a magnetic field of 0, 4 and 7~T$\equiv$
6.6~cm$^{-1}$. We neither observe a shift nor a splitting nor even
a broadening of the 67- cm$^{-1}$~mode. Therefore such a simple
one-magnon interpretation can be clearly ruled out.

On the other hand, one might also suggest that our three bound
states are not of magnetic but rather of phononic origin. Indeed,
such phonon bound state phenomena have been observed for example
in YbS \cite{merlin78,vitins}. There, these modes were interpreted
as excitons interacting with an LO phonon of frequency
$\omega_{0}$ giving rise to exciton-phonon bound states with
frequencies $\omega$ = {\it n} $\times$ $\omega_{0}$, with n an
integer and polarization selection rules identical with the
original LO phonon. Finally, the linewidth of these bound states
is a linear function of {\it n}. The first property can be found
in our system on assuming an $\omega_{0}$ of about 30~cm$^{-1}$.
However, no such phonon mode could be detected. Besides, the
scattering intensity of the bound states in
$\alpha^\prime$-NaV$_{2}$O$_{5}$ appears not only in the fully
symmetric scattering components (aa), (bb), but also in (ab). A
similar disagreement is found for the linewidth that does not show
a systematic broadening with n. Clearly, these experimental
results are not compatible with our observations in
$\alpha^\prime$-NaV$_{2}$O$_{5}$.

\begin{figure}
\centerline{\psfig{file=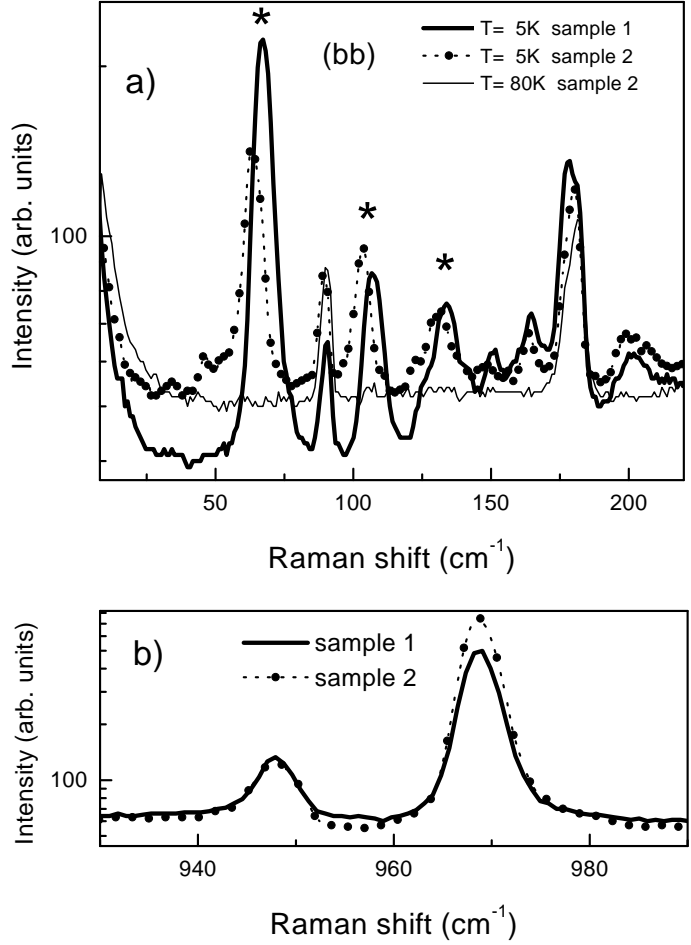,height=15cm,rheight=15cm}}
\caption{Comparison between sample 1 and the 1 \% Na deficient
sample 2 of $\alpha^\prime$-NaV$_{2}$O$_{5}$ at T=5~K.
Additionally a measurement of sample 2 at T=80~K is given in a).
The scattering intensity is displayed on a logarithmic scale. The
magnetic bound states in a) marked with an asterisk show a
considerable shift toward lower frequencies with Na deficiency
while the modes at 202 cm$^{-1}$ in a) and 948 cm$^{-1}$ in b)
stay constant in frequency. Please note that the decrease of the
background scattering intensity of sample 1 for frequencies below
2$\Delta_{ST}$=120~cm$^{-1}$ is suppressed by the Na deficiency of
sample 2.} \label{f5}
\end{figure}

We will now use results on the Na deficient sample 2 to confirm
the above made analysis and to prove that the energy of the bound
states is closely related to the singlet-triplet gap. One of the
most subtle problems preparing $\alpha\prime$-NaV$_{2}$O$_{5}$
samples is to control the Na content. Deviations from the nominal
stoichiometry result in a shift of the ratio between V$^{4+}$ and
V$^{5+}$ toward the nonmagnetic (S=0) V$^{5+}$. This then acts as
an effective substitution. Systematic studies of Na deficiencies
as reported in Ref. \onlinecite{isobe98} indeed show that the drop
in the magnetic susceptibility is suppressed and T$_{\rm SP}$ is
slightly shifted toward lower temperatures. Comparing the data in
Ref. \onlinecite{isobe98} with the susceptibility data of our
sample 2 we come to the conclusion that the latter has about 1\%
Na deficiency. In principle, both samples show the same Raman
scattering results both in the high temperature and the dimerized
phase. In the dimerized phase the frequencies of most
transition-induced modes are the same for both samples within the
experimental resolution. Solely, the three low-frequency modes
show a considerable shift in frequency to 64 (67), 103 (107), and
130 (134) cm$^{-1}$, with the values of sample 1 given in
brackets. A direct comparison of the T=5~K spectra of sample 1 and
2 in (bb)-polarization is shown in Fig. \ref{f5}. Clearly, the
modes at 202, 246 (not shown), and 948~cm$^{-1}$ attributed to
zone folded phonon modes do not show this shift towards lower
frequencies. The frequency shift of the bound states is explained
quite naturally by assuming that the Na deficiency of sample 2
leads to a reduced singlet-triplet gap and hence to a smaller
energy of the magnetic bound states. Actually, a similar behavior
has been observed in Zn-substituted CuGeO$_{3}$ \cite{lemmens}.

\begin{figure}
\centerline{\psfig{file=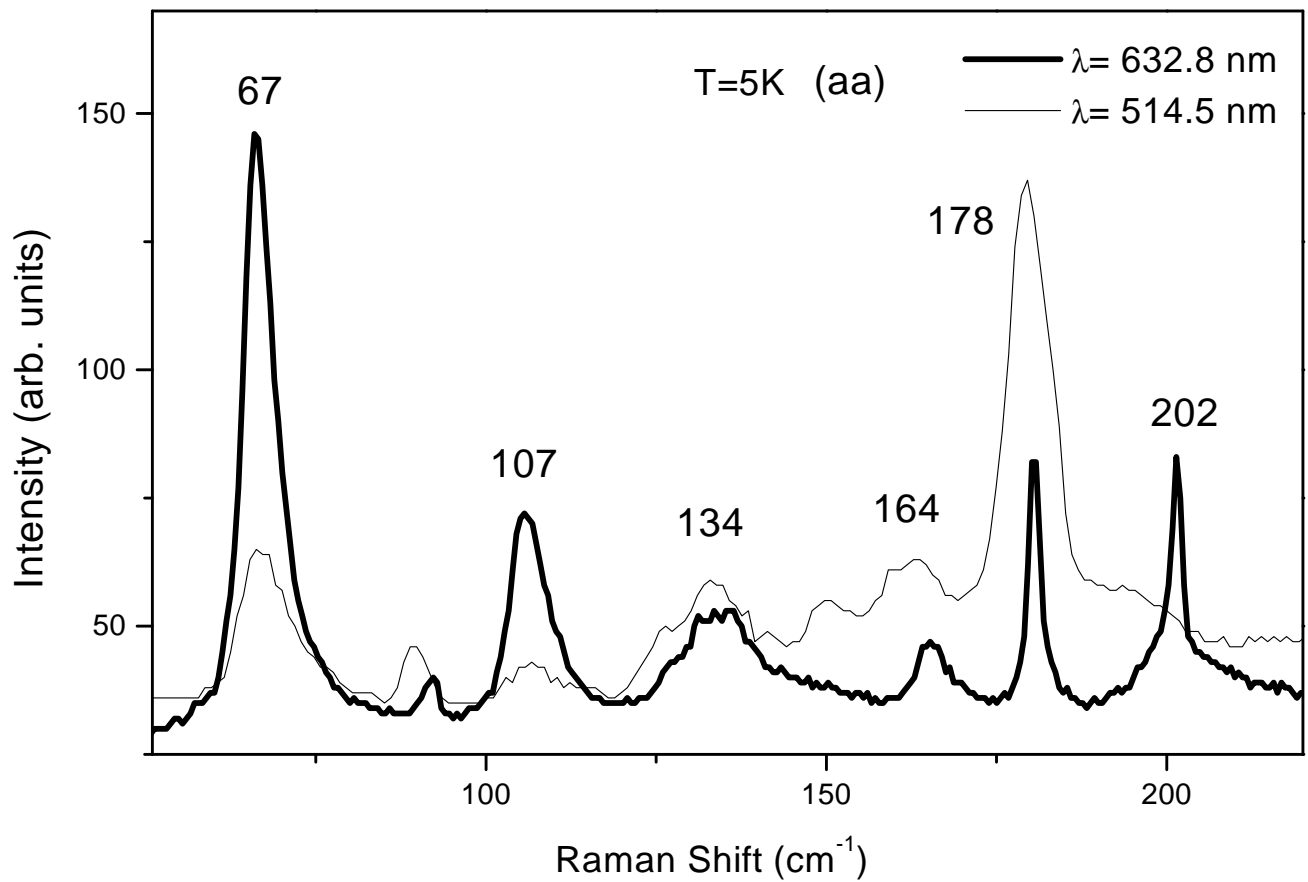,height=14cm,rheight=14cm}}
\caption{Comparison of the (aa) spectra of
$\alpha^\prime$-NaV$_{2}$O$_{5}$ at 5~K for $\lambda$=514.5nm and
$\lambda$=632.8~nm.} \label{f9}
\end{figure}

Further insight into the underlying physics of the low temperature
excitations can be gained from resonance Raman measurements. By
changing the frequency of the incident laser light from
$\lambda$=514.5~nm (2.5~eV) to $\lambda$=632.8~nm (1.9~eV) one
observes a different response of the phonon intensity and the
three magnetic bound states. Fig.\ref{f9} shows the low frequency
part of the (aa) spectrum at 5~K for $\lambda$=514.5~nm and
$\lambda$=632.8~nm. It can be clearly stated that compared to both
the high temperature mode at 178~cm$^{-1}$ and the folded phonon
mode at 164~cm$^{-1}$ the three magnetic bound states at 67, 107
and 134~cm$^{-1}$ gain intensity for excitation at smaller
wavelength. This again underlines our above made distinction
between modes of phononic and modes of magnetic origin as the
magnetic excitations are coupled differently to the optical
excitation.

The dependence of the gained scattering intensity on the frequency
of the incident laser light can easily be understood in the way
that there are different electronic transitions involved as
intermediate states in the Raman process. Roughly speaking, the
scattering intensity I in a two-band approximation should behave
like I~$\sim$~1/($\omega$-$\Delta_{\rm et}$)$^{p}$, with $\omega$
the frequency of the incident laser light and $\Delta_{\rm et}$
the energy of the electronic transition involved. The parameter
$p$ depends on the order of the scattering process.


\subsection{Ground state properties of $\alpha^\prime$-NaV$_{2}$O$_{5}$}
It can be stated that the modes at 67, 107 and 134~cm$^{-1}$ can
be described best as magnetic singlet bound states. They, however,
do not fit both in number and in selection rules to the excitation
spectrum of a frustrated and dimerized 1D Heisenberg chain, which
exhibits only one singlet bound state. The intensity of this
singlet bound state should only be observed with polarizations
parallel to the dominant exchange path, i.e. along the chain
direction.

In the high temperature phase (T$\ge$ T$_{\rm SP}$)
$\alpha^\prime$-NaV$_{2}$O$_{5}$ is described as a quarter-filled
Hubbard ladder. As the charge transfer gap
$\Delta_{CT}\approx$1~eV is large the low energy excitations are
dominated by spin fluctuations of one spin per rung along the
ladder direction. This leads to a mapping of the system on a spin
1/2 Heisenberg chain \cite{smolinski}. Actually, the mode at 107
cm$^{-1}$ does match both in selection rules (only observable with
polarization of the incoming and scattered light along the ladder
direction) and in energy (107cm$^{-1}$=1.78$\times$ $\Delta_{\rm
ST}$) the singlet bound state in CuGeO$_{3}$ (30
cm$^{-1}$=1.76$\times$ $\Delta_{\rm ST}$). Therefore we can
understand this mode as the singlet bound state of the effective
dimerized chain. However, there are two further bound states at 67
and 134 cm$^{-1}$ which are unpolarized in the ab plane.

This breaking of the selection rules expected for an effective
chain system may be understood if the full geometry of the
Heisenberg ladder is taken into account. This means that both
dimers along the ladder diagonal as well as dimers along and
perpendicular to the ladder exist. As the observed bound states
have a small energy separation compared to
$\Delta_{CT}\approx$1~eV the underlying dimers should be
energetically nearly degenerate. Actually, their energy scale is
given by the singlet-triplet gap $\Delta_{ST}\approx$9.8~meV
\cite{yoshihama}. Therefore the character of the excitations is
predominantly magnetic allowing a description of the bound states
in the context of a pure spin model.

Furthermore, as the bound states are unpolarized no preferred
dimer configuration exists. The small energy separation therefore
allows for a dynamic dimer formation as the new ground state of
this system. In this way one can both explain the unexpected
multiplicity and the breakdown of the selection rules
\cite{lemmens98}. There exists a close analogy of this picture to
the dynamical Jahn-Teller or the RVB model. The superposition of
different energetically nearly degenerate dimer configurations
should also show up in the triplet channel of the excitation
spectrum. In fact, recent neutron scattering studies
\cite{yoshihama} revealed an unexpected splitting of the
dispersion of the triplet excitations along the a-axis. This is
consistent with our considerations as there are dimers not only in
chain direction but also perpendicular to it. Further theoretical
support of our point of view comes from two recent works
\cite{seo,mostovoy} assuming a zig-zac chain structure
corresponding to diagonal dimers as the ground state for
T$<$T$_{\rm SP}$. Due to charge ordering driven by
intersite-Coulomb interaction such a zig-zag type of charge
disproportionation along the ladders occurs leading to dimers
which are not only situated along the chain direction.

\subsection{Charge ordering}

Despite the above explanation of the low energy Raman spectrum in
terms of a spin model with resonating dimer configurations the
underlying mechanism that drives the system into the dimerized
ground state is nevertheless still under discussion. One of the
key questions is whether a charge ordering occurs and whether its
energy scale is constrained by the large $\Delta_{CT}$.
Furthermore, it should be clarified whether charge ordering
appears instantaneously with the singlet formation or whether two
distinct consecutive transitions can be observed.

Experimental evidence for a charge ordering is given by the
observation of two distinctly different V sites below T$_{{\rm
SP}}$ in NMR spectroscopy \cite{ohama98}. In thermal expansion
data a double phase transition has been observed in a very small
temperature interval near T=34~K \cite{koeppen}. However, the
interpretation of the data is not unambiguous. On the other hand,
recent infrared measurements point towards a temperature scale
different from T$_{\rm SP}$ because of the finite intensity of a
spin-Peierls active phonon for temperatures as high as 60-70~K
\cite{damanavo}. Recent ultrasonic experiments \cite{fertey}
report a transition induced anomaly in the longitudinal sound
velocity along the chain direction extending up to 2 $\times$
T$_{\rm SP}\approx$ 70~K. On the low temperature side the
transition appears to be much sharper. Indeed, the velocity
anomaly disappears again only at 0.8 $\times$ T$_{\rm SP}$=27~K.
This corresponds to the temperature at which the magnetic modes
start to develop in the Raman spectra supporting their
interpretation as magnetic bound states that need to have a well
developed order parameter.

\begin{figure}
\centerline{\psfig{file=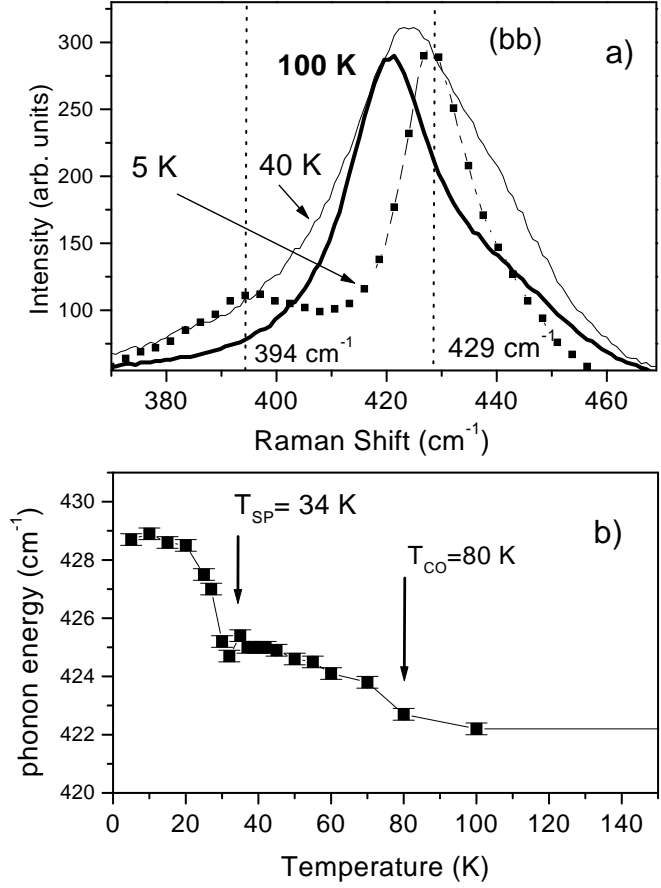,height=14cm,rheight=14cm}}
\caption{Temperature dependence of the energy of the 422-cm$^{-1}$
mode of $\alpha^{\prime}$-NaV$_{2}$O$_{5}$ in (bb) polarization:
a) Scattering intensity and b) phonon energy. For temperatures
below 22K a second peak develops near 394~cm$^{-1}$. The
temperature dependence of the mode at 422~cm$^{-1}$ marks the
onset of charge ordering at T$_{CO}$=80~K.} \label{f2}
\end{figure}

At room temperature all the V sites have formally the valence
+4.5. In this case a $z-$axis vibration (out of the ab plane) of
the V ions within one rung should be Raman-silent if the ions move
out-of-phase. At the same time, the corresponding in-phase
vibration is Raman active. A charge disproportionation within a
rung can manifest itself in Raman scattering as the appearance of
a new intense mode. This mode is related to V ion vibrations
accompanied by a considerable shift in frequency of other
V-related modes. This shift occurs mainly due to  ordering-induced
changes in the Coulomb contribution to the lattice force
constants. Experimentally, we observe such a behavior of a phonon
in A$_{g}$ symmetry at 422~cm$^{-1}$ that shifts to 429~cm$^{-1}$
at lower temperatures with an additional split-off mode at
394~cm$^{-1}$. Fig. \ref{f2} shows these modes in (bb) scattering
geometry, which we suppose to be the in-phase $z-$axis vibration
of the V ions within a rung. In the case of charge asymmetry, the
normal coordinates are linear combinations of the in-phase and
out-of-phase vibrations, and, therefore a new ''in-phase'' mode
appears in the Raman spectrum. If the mixing is strong enough,
each one of the new modes can be interpreted as a vibration of one
V ion. Below T$_{{\rm SP}}$ the frequency of the intense peak
shifts to 429~cm$^{-1}$ at 5~K. Comparing the temperature
dependencies of the frequency and the intensity of this additional
phonon below T$_{{\rm SP}}$ it can be stated that the onset of the
hardening is well above this temperature.

The amount of hardening can roughly be estimated quantitatively.
We define the Coulomb contribution to the force constant $k_{{\rm
Coulomb}}$ as the second derivative of the Coulomb energy with
respect to the $z-$displacement of the V ion for its interaction
with the apical oxygen above the V ion and the nearest-neighbors
in the ($ab$) plane. Thereby we get $k_{{\rm Coulomb}}$~$\approx
$~2.5$\times 10^{5}$~erg/cm$^{2}$ \cite{sherman}. The measured
effective $k$ for this mode can be estimated by
$k_{\text{exp}.}$=$M$$\Omega ^{2}$ with $M$ the V ion mass and
$\Omega $=420~cm$^{-1}$, to $k_{\text{exp}.}$$\approx $~6$\times
$10$^{5}$~erg/cm$^{2}$. The difference $k_{\text{exp}.}$-$k_{\rm
Coulomb}$ should be attributed to a further contribution due to
the covalent bonding $k_{\exp.}$=$k_{{\rm Coulomb}}$+$k_{{\rm
bonding}}$. This contribution of $k_{\rm bonding}$ between a V ion
and the apical O should not be too much affected by a charge
ordering. The charge asymmetry, on the other hand, changes the
charge of one of the V ions from +4.5 to +5, i.e.
$\widetilde{k}_{{\rm Coulomb}}$=$k_{{\rm Coulomb}}\times $1.1.
With this value one can estimate the new frequency
$\widetilde{\Omega}=\sqrt{\widetilde{k}/M}$. The corresponding
shift $\widetilde{\Omega}-\Omega $ that should be observable as
function of temperature is then 2 \% or $\approx $ 8~cm$^{-1}$.
This corresponds roughly to the shift of the mode from
422~cm$^{-1}$ to 429~cm$^{-1}$.

A new peak appears in the dimerized phase at 394~cm$^{-1}$ as a
shoulder of the 429~cm$^{-1}$ mode. We attribute this peak to the
vibration originating from the silent mode. It then can be
assigned to arising from the scattering mainly by V$^{4+}$
vibrations, whereas the peak at 429~cm$^{-1}$ is due to the
V$^{5+}$ vibrations. The splitting of the frequencies is due to a
direct interaction of the V ions. The diminished intensity of the
peak at 394~cm$^{-1}$ compared to the 429-cm$^{-1}$ mode can
qualitatively be understood by taking into account that V$^{4+}$
possesses one more electron compared to V$^{5+}$. Therefore, a
virtual transition of an electron to this ion requires an
additional energy due to the Coulomb repulsion with the
$d$-electron already there. If the photon energy is out of
resonance with this transition, this Hubbard-like term will
diminish the scattering intensity of the V$^{4+}$ vibration at
394~cm$^{-1}$.

It is worth mentioning that the charge ordering can hardly be
considered as static since the phonons can drive strong
fluctuations of the V ion charge. Let us consider the situation in
more details at the beginning for T$>$T$_{\text{SP}}$. The
$d-$electrons are in the strong Coulomb field of the apical oxygen
above the V ion. In the out-of-phase vibration the V ions become
nonequivalent, and the field causes strong charge transfer between
them. The relative value of the charge transfer $\Delta
Q/\overline{Q}$ can be estimated as $eE_{\rm a}z_{0}/t_{\bot },$
where $e$ is the electron charge, $E_{\rm a}$ is the electric
field between the apical oxygen and the V ion, $z_{0}$ is the
zero-point vibrational amplitude, $t_{\bot }$ is the $a
$-direction hopping perpendicular to the ladder, and
$\overline{Q}=1/2$. Taking as estimate $E_{a}\approx
2e/d_{0}^{2}\approx 4\times 10^{6}$ cgs units, with $d_{0}\approx
1.6 \AA$ the distance between the V and apical oxygen ions, the
zero-point vibrational amplitude $z_{0}=\sqrt{\hbar /2M\Omega
}\approx 0.03~\AA$, with $\Omega \approx 422$ cm$^{-1}$, and
$t_{\bot }\approx 0.35$ eV \cite{horsch}, we obtain $\Delta
Q/\overline{Q}\sim 1$. This implies very strong phonon-driven
charge fluctuations \cite{sherman}. Since the superexchange is
determined by the hopping of electrons between the rungs, these
charge fluctuations, in turn, cause fluctuations of the exchange
parameter $J,$ leading to a considerable spin-phonon coupling
\cite{sherman2,sherman3}. In the low-temperature phase the charge
fluctuations are smaller than above T$_{\text{SP}}.$ However, they
are still strong enough to make the ordering non-static. Therefore
the fluctuations have to be considered as one of the driving
forces for dynamical dimer configurations as manifested in the
multiplicity of the observed magnetic bound states.

The temperature dependent frequency shift of the 422 cm$^{-1}$
mode displayed in Fig. \ref{f2} shows that this charge ordering
partly exists above T$_{{\rm SP}}$. At these temperatures the
ordering can occur as zigzag-like fluctuations of V ions charges
in parts of the ladders. Therefore a crossover temperature of
T$_{\rm CO}$=80~K is defined as the onset of the frequency shift
compared to higher temperatures. This smooth crossover is typical
for charge ordering in 1D. A similar behavior has been observed in
neutron diffraction studies of stripe ordering in
La$_{2}$NiO$_{4+\delta}$ \cite{wochner}. The super structure peaks
corresponding to stripes of charges and of spins appear in this
compound separated in temperature for T$\leq$T$_{CO}$=200~K and
T$\leq$T$_{m}$=110~K, respectively. The appearance of T$_{CO}$ in
the temperature dependence of the superlattice intensity is
identical to our observations in $\alpha^\prime$-NaV$_{2}$O$_{5}$.
As T$_{CO}$=200~K exceeds T$_{m}$ by a factor of 2 charge ordering
is proposed to be the driving force of this phenomenon in
La$_{2}$NiO$_{4+\delta}$ \cite{lee97}. The same conclusion should
hold for $\alpha^\prime$-NaV$_{2}$O$_{5}$ as here charge ordering
appearing below T$_{CO}$=80~K sets the scale for the following
singlet formation at T=34~K.

\subsection{Fluctuations above the singlet ground state formation}
\begin{figure}
\centerline{\psfig{file=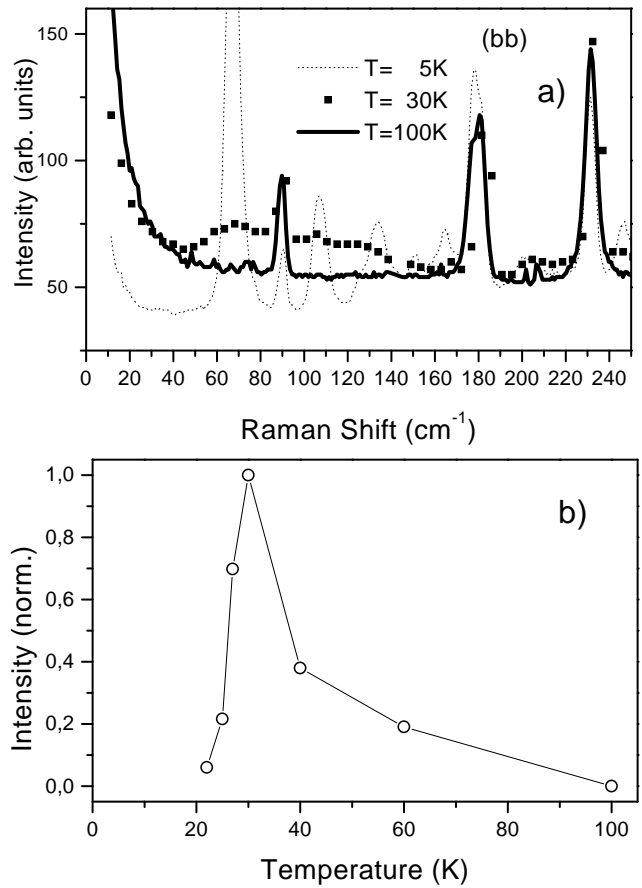,height=14cm,rheight=14cm}}
\caption{ a) Additional scattering contribution of
$\alpha^{\prime}$-NaV$_{2}$O$_{5}$ at 30~K in the low frequency
range (40-160~cm$^{-1}$) in (bb) polarization and b) temperature
dependence of its integrated intensity.} \label{f6}
\end{figure}

Now the question arises whether the charge ordering might also
show up in fluctuations in the magnetic light scattering. We will
present two further low frequency scattering contributions that
are detected for temperatures below T$_{CO}$ and assigned in that
way. In Fig.\ref{f6} the low frequency part of the (bb) spectrum
is presented for T=5~K, 30~K, and 100~K. At 100~K only the high
temperature phonons at 90, 178 and 223~cm$^{-1}$ are observed. For
T=5~K this spectrum is modified due to the existence of the
additional low temperature modes. On the contrary, for T=30~K one
finds a broad scattering contribution from 40 to 160~cm$^{-1}$.
This scattering contribution arises for temperatures below 80~K
and peaks in intensity at about 30~K (Fig. 8b)). With the low
frequency modes emerging below T$_{\rm SP}$ the intensity of this
broad contribution drops again. There is this clear competition
between the broad low frequency scattering contribution and the
magnetic bound states both in their intensity as function of
temperature and in the common frequency range. Therefore, it
appears straightforward to assign this contribution to  magnetic
light scattering that is suppressed for T $\le$ T$_{\rm SP}$.

\begin{figure}
\centerline{\psfig{file=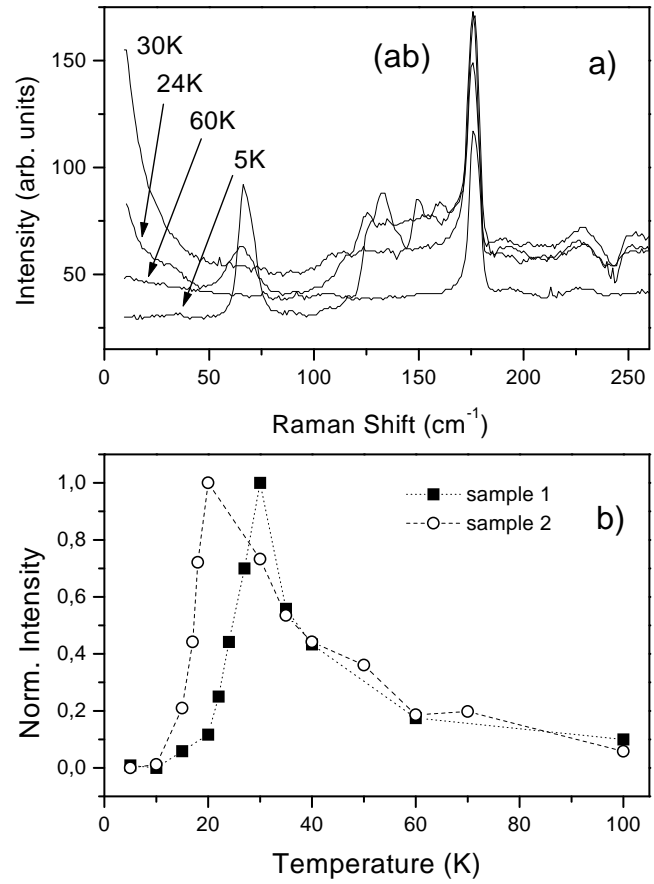,height=14cm,rheight=14cm}}
\caption{Quasielastically scattered light of
$\alpha^{\prime}$-NaV$_{2}$O$_{5}$ below T$_{\rm CO}$ observed in
(ab) polarization. a) Spectra of sample 1 for different
temperatures. b) Temperature dependence of the normalized
integrated intensity comparing sample 1 and sample 2.} \label{f7}
\end{figure}

A second observation of pretransitional fluctuations exists in the
low frequency data in (ab) scattering geometry. The Raman data in
this geometry is presented as a function of temperature in
Fig.\ref{f7} a). A quasielastic scattering contribution is
observed which again increases below 80~K peaking near T$_{\rm
SP}$. Again this contribution is suppressed at lower temperatures
due to the appearance of the magnetic bound states. In
Fig.\ref{f7} b) the temperature dependence of this intensity is
shown. Clearly, it exhibits the same behavior as the broad
contribution in (bb) polarization (Fig. 8b)). For comparison we
have included the temperature dependence of the sample 2 which has
a 1\% Na deficiency. Here the transition region appears to be
broadened and the maximum signal occurs at lower temperatures
compared to sample 1. However, it is important to note that the
drop in this quasielastic light scattering contribution light does
not coincide with T$_{\rm SP}$ but rather with the occurrence of
the magnetic bound states. These states emerge at considerable
lower temperatures compared to the zone-folded phonons. Therefore
this quasielastic scattering contribution can be clearly
identified as originating from magnetic Raman scattering. In this
sense the $\omega~\sim~0$ scattering as well as the broad
contribution in (bb) scattering geometry can be understood as
dimer fluctuations in a charge ordered state. The main difference
between both effects is that the $\omega~\sim~0$ scattering in
(ab) polarization is gapless.

\begin{figure}
\centerline{\psfig{file=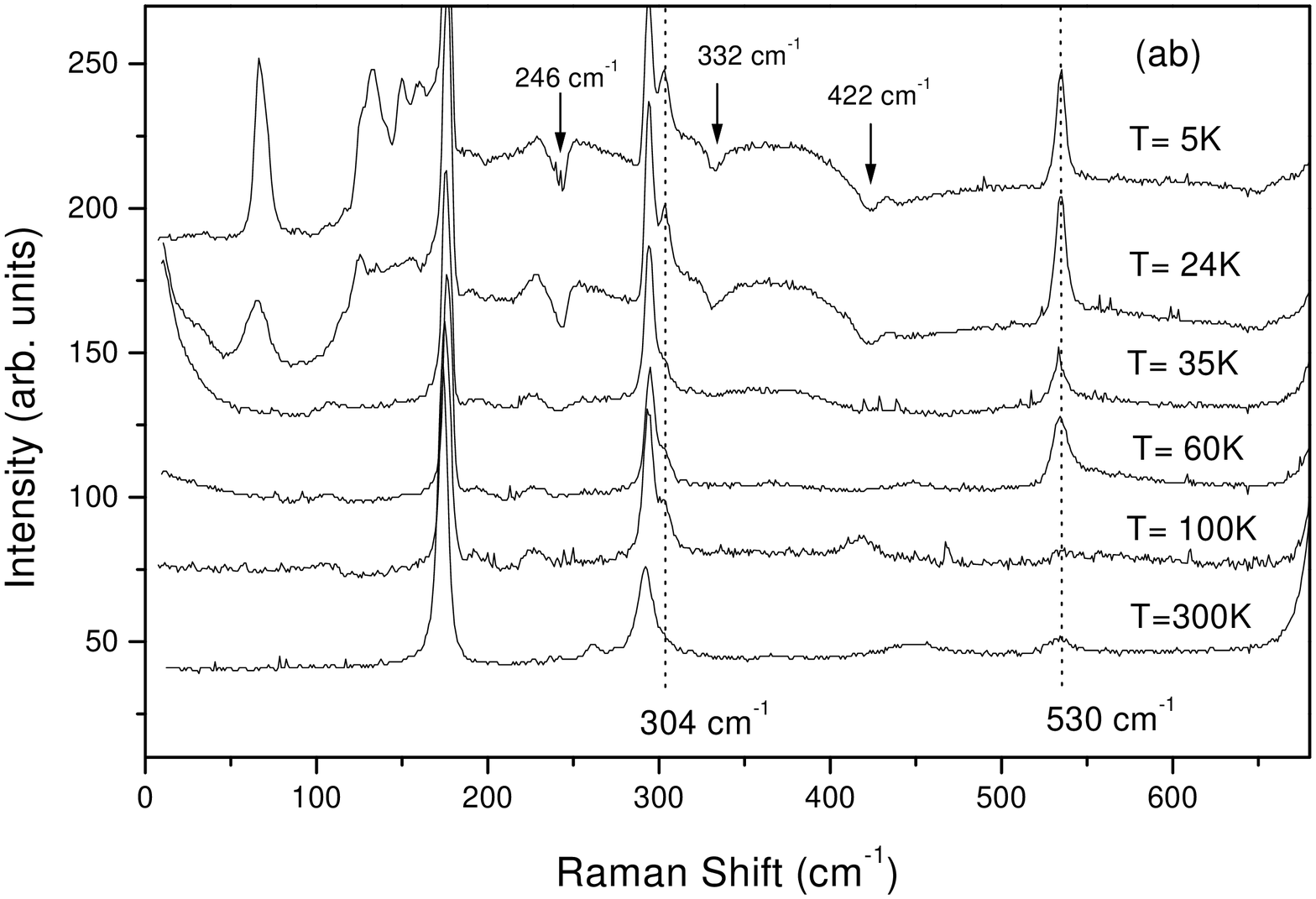,height=14cm,rheight=14cm}}
\caption{ Raman spectra of $\alpha^{\prime}$-NaV$_{2}$O$_{5}$ in
(ab) polarization for several temperatures. For clarity the
spectra are shifted against each other. For T$<$60K symmetry
forbidden A$_{1g}$ phonons evolve at 304 and 530~cm$^{-1}$. Below
T$_{\rm SP}$ there are three dips (antiresonances) in the
scattering continuum at 246, 332, and 422~cm$^{-1}$ which coincide
in frequencies with phonon modes observed in (bb) polarization;
the first two modes are folded zone boundary phonons for
T$<$T$_{SP}$.} \label{f8}
\end{figure}

A further phonon anomaly on the temperature scale of T$_{\rm
CO}$=80~K is displayed in Fig.\ref{f8}. For the centrosymmetric
space group D$_{2h}^{13}$ the (ab)-spectrum should show only
B$_{1g}$ modes. For the room temperature spectrum this is actually
the case. There are essentially only three modes at 173, 292, and
692 cm$^{-1}$ (not shown) in agreement with the factor group
analysis. However, already for temperatures below $\sim$100~K a
significant contribution of the A$_{1g}$ mode at 530 cm$^{-1}$ can
be detected. Also the 304 cm$^{-1}$ mode gradually appears. Such
an appearance of symmetry forbidden A$_{1g}$ modes can be
understood presuming a breakdown of the inversion symmetry due to
charge ordering at this temperature. Actually, in the (ab)
spectrum at 5~K one can see not only the above mentioned A$_{1g}$
modes but furthermore also some dips in the broad continuum at
246, 332, and 422~cm$^{-1}$. These antiresonances origininate from
phonons observed in the (bb) spectrum. This points to an
interaction between these phonons and the two-triplet continuum in
(ab) polarization. Finally, the B$_{1g}$ mode at 692 cm$^{-1}$ is
also visible in the (aa) spectrum in the dimerized phase. Thus the
charge ordering does not only show up in the magnetic scattering
but also in the lattice degrees of freedom.

\section{CONCLUSIONS}
We have performed Raman scattering studies on
$\alpha^\prime$-NaV$_{2}$O$_{5}$. Room temperature phonon
scattering is in agreement with a crystal structure that leads to
a quarter-filled ladder as the appropriate underlying physical
model. The transition-induced modes (T~$\le$~T$_{\rm SP}$) are
divided into magnetic bound states at 67, 107, 134~cm$^{-1}$ and
zone boundary folded phonons. The properties of the bound states
lead us to the conclusion that the ground state of this system can
be understood as a superposition of various nearly degenerate
dimer configurations. As the energy scale of the bound states is
comparable to the singlet triplet gap their predominantly magnetic
character is concluded, allowing a discussion within purely
magnetic spin models. The phase transition into this ground state
is initiated by a charge ordering of the V$^{4+}$/V$^{5+}$ ions
occurring as a smooth crossover at temperatures higher than
T$_{\rm SP}$. Experimental evidence for such a charge ordering is
found as an anomalous hardening and splitting of a V vibration at
422~cm$^{-1}$. This hardening sets in at a  crossover temperature
T$_{\rm CO}$=80~K and is accompanied by fluctuations which were
identified in a quasielastic scattering in (ab) scattering
geometry and a broad continuum extending from 40~cm$^{-1}$ to
160~cm$^{-1}$ in the (bb) scattering contribution. The onset of a
charge ordering is further supported by a breakdown of phonon
selection rules for T$\le$100~K. The subtle interplay of charge
and spin degrees of freedom manifests itself in
$\alpha^\prime$-NaV$_{2}$O$_{5}$ in a transition into a dimerised
ground state at T$_{\rm SP}$=34~K which is initiated by a charge
ordering for temperatures below T$_{\rm CO}$=80~K, as well as in a
rich excitation spectrum. Therefore the concluded similarities
with the behavior of La$_{2}$NiO$_{4+\delta}$ or quasi-one
dimensional conductors are far from being accidental. These
partially filled low-dimensional systems are all governed by
Coulomb correlations.


Acknowledgment: We acknowledge stimulating discussions with G.
Uhrig, C. Gros, W. Brenig, B. B\"uchner, M. Konstantinovic, M.
Grove, H. Seo, H. Fukuyama, T. Yosihama, K. Nakajima, K. Kakurai,
and M. Udagawa. E.Y.S. acknowledges partial support from the
Humboldt foundation. Furthermore support by DFG through SFB 341
and SFB 252, by INTAS 96-410 and by BMBF Fkz 13N6586/8 is
gratefully acknowledged.

\begin{table}
\caption{Phonon frequencies of $\alpha^{\prime}$-NaV$_{2}$O$_{5}$
at room temperature (T=300~K) in the scattering geometries
accessible. For comparison the expected number of modes from the
factor group analysis for the space groups C$_{2v}^{7}$ and
D$_{2h}^{13}$ are given.}
\begin{center}
\begin{tabular}{|c|ccccccccccc|c|c}
Polarization&\multicolumn{11}{c|}{Phonon frequencies (cm$^{-1}$)}
&(C$_{2v}^{7}$)&(D$_{2h}^{13}$)
\\\tableline (aa)&90&&178&&&304&422&450&530&&970&&\\
(bb)&90&&178&230&&304&422&450&530&&&15&8\\
(cc)&90&&178&&&&422&&&&970&& \\\hline
 (ab)&&173&&&292&&&&&692&&7&3
\\
\end{tabular}
\end{center}
\end{table}
\begin{table}
\caption{Additional low temperature modes of
$\alpha^{\prime}$-NaV$_{2}$O$_{5}$ for T$\le$T$_{\rm SP}$ in
various scattering geometries. For comparison the frequencies of
the modes in sample 2 are given in brackets.}
\begin{center}
\begin{tabular}{|c|cccccc}
Polarization&\multicolumn{6}{c}{Low temperature modes (cm$^{-1}$)
for T$\le$T$_{\rm SP}$}
\\\tableline
(aa)&67(64)&&133(130)&&&\\
(bb)&67(64)&107(103)&134(131)&151&164(164)&202(202)\\
(ab)&67(64)&&134(131)&&& \\\hline
(aa)&&&&650(650)&692(692)&948(947)\\
(bb)&246(244)&325(325)&396(396)&&&948(948)\\ (ab)&&&&&&948(948)\\
\end{tabular}
\end{center}
\end{table}

\begin{thebibliography}{99}

\bibitem{pytte} E. Pytte, Phys. Rev. {\bf B10}, 4637 (1974).

\bibitem{cross} M.C. Cross and D.S. Fisher, Phys. Rev. {\bf B19}, 402 (1979).

\bibitem{bray}For a review, see J.W. Bray, L.V. Interrante, I.S. Jacobs, and J.C. Bonner,\
in {\it Extended Linear Chain Compounds,}
J.S. Miller (Ed.), Plenum, New York, 1983, Vol. 3, pp. 353-415.

\bibitem{hase} M. Hase, I. Terasaki, K. Uchinokura, Rev. Lett. {\bf 70}, 3651 (1993).

\bibitem{nishi} M. Nishi, O. Fujita, and J. Akimitsu, Phys. Rev. {\bf B50},
6508 (1994).

\bibitem{emery} G. Castilla, S. Chakravarty, and V.J. Emery, Phys. Rev.
Lett. {\bf 75}, 1823 (1995).

\bibitem{boucher} For a review, see J.P.
Boucher and L.P. Regnault, J. Phys I (France) {\bf 6}, 1 (1996).

\bibitem{kuroe} H. Kuroe, T. Sekine, M. Hase, Y. Sasago, K. Uchinokura,
H. Kojima, I. Tanaka, and Y. Shibuya, Phys. Rev. {\bf B50}, 16468
(1994).

\bibitem{mutu} V.N. Muthukumar, C. Gros, W. Wenzel, R. Valenti, P.Lemmens,
B. Eisener, G. G\"untherodt, M. Weiden, C. Geibel, F. Steglich,
Phys. Rev. {\bf B54}, R9635 (1996).

\bibitem{lemmens} P. Lemmens, M. Fischer, G. G\"untherodt, C. Gros, P.G.J. van Dongen,
M. Weiden, W. Richter, C. Geibel, and F. Steglich, Phys. Rev. {\bf
B55}, 15076 (1997).

\bibitem{kotov98} V.N. Kotov, O.P. Sushkov and R. Eder,
cond-mat/9808169.


\bibitem{isobe} M. Isobe and Y. Ueda, J. Phys. Soc. Jpn. {\bf 65}, 1178
(1996).

\bibitem{weiden} M. Weiden, R. Hauptmann, C. Geibel, F. Steglich, M. Fischer,
P. Lemmens, G. G\"untherodt, Z. Phys. {\bf B103}, 1 (1997).

\bibitem{fuji} Y. Fujii, H. Nakao, T. Yosihama, M. Nishi, K. Nakajima, K. Kakurai,
M. Isobe, Y. Ueda, H. Sawa, J. Phys. Soc. Jpn. {\bf 66}, 326
(1997).

\bibitem{galy} J. Galy, A. Casalot, M. Pouchard, P. Hagenmuller,
C. R. Acad. Sc. Paris, {\bf C262}, 1055 (1966).

\bibitem{roth} H.G.v. Schnering, Yu. Grin, M. Kaupp, M. Somer, R.K. Kremer,
O. Jepsen, T. Chatterji, and M. Weiden, Z. f. Kristallogr.-New
Cryst. Struct. {\bf 213}, p. 246 (1998).

\bibitem{smolinski} H. Smolinski, C. Gros, W. Weber, U. Peuchert, G. Roth, M. Weiden,
and C. Geibel, Phys. Rev. Lett. {\bf 80}, 5164 (1998).

\bibitem{horsch} P. Horsch and F. Mack, Eur. Phys. J. {\bf B 5}, 367 (1998).

\bibitem{seo97} H. Seo and H. Fukuyama, Journ. Phys. Soc. Jpn. {\bf
66}, 1249 (1997).

\bibitem{thalmeier} P. Thalmeier and P. Fulde, Europhys. Lett. {\bf 44}, 242
(1998).

\bibitem{mostovoy} M.V. Mostovoy, and D.I. Khomskii, cond-mat/9806215, preprint (unpublished).

\bibitem{seo} H. Seo, H. Fukuyama, cond-mat/9805185, preprint (unpublished).

\bibitem{yoshihama} T. Yoshihama, M. Nishi, K. Nakajima, K.
Kakurai, Y. Fujii, M. Isobe, C. Kagami and Y. Ueda, Journ. Phys.
Soc. Jpn. {\bf 67}, 744 (1998).

\bibitem{gros} C. Gros, and R.
Valent\'{i}, Phys. Rev. Lett. {\bf 82}, 976 (1999).

\bibitem{rousseau} D.L. Rousseau, R.P. Bauman, and S.P.S. Porto, Journal of Raman Spectroscopy,
{\bf 10}, 253 (1981).

\bibitem{damanavo} A. Damascelli, D. van der Marel, M. Gr\"uninger, C. Presura, T.T.M. Palstra,
J. Jegoudez, and A. Revcolevschi, Phys. Rev. Lett. {\bf 81}, 922
(1998).

\bibitem{golub} S.A. Golubschik, M. Isobe, A.N. Ivlev, B.N. Mavrin, M.N. Popova,
A.B. Sushkov, Y. Ueda, A.N. Vasilev, J. Phys. Soc. Jpn. {\bf 66},
4042 (1997).


\bibitem{yamada} I. Yamada and H. Onda, Phys. Rev. {\bf B49}, 1049
(1994).

\bibitem{kuroea} H. Kuroe, J. Sasaki, T. Sekine, N. Koide, Y. Sasago, K. Uchinokura,
M. Hase, Phys. Rev. {\bf B55}, 409 (1997).


\bibitem{brenig} W. Brenig, Phys. Rev. {\bf B56}, 2551 (1997).

\bibitem{lockw} D.J. Lockwood, in Light Scattering in Solids III, Topics in
Appl. Physics, Vol 51, ed. by M.~Cardona and G.~G\"untherodt,
Springer Verlag Berlin 1982, p. 59 and references therein.

\bibitem{misochko} O.V. Misochko, E.Ya. Sherman, Physica {\bf C222}, 219 (1994).

\bibitem{lemmens98} P. Lemmens, M. Fischer, G. Els, G.
G\"untherodt, A.S. Mishchenko, M. Weiden, R. Hauptmann, C. Geibel,
and F. Steglich, Phys. Rev. {\bf B58}, 14159 (1998).

\bibitem{bouz} G. Bouzerar, A.P. Kampf, and G.I. Japaridze,
Phys. Rev. {\bf B 58}, 3117 (1998).

\bibitem{schmidt98} S. Schmidt, W. Palme, B. L\"uthi, M. Weiden, R.
Hauptmann, and C. Geibel, Phys. Rev. {\bf B57}, 2687 (1998).

\bibitem{kuroe98b} H. Kuroe, H. Seto, J. Sasaki, T. Sekine, M.
Isobe, and Y. Ueda, Jour. Phys. Soc. Jpn. {\bf 67}, 2881 (1998).

\bibitem{cottam}M. Cottam and D. Lockwood, in Light Scattering in
Magnetic Solids, John Wiley New York 1986, p. 135 ff.

\bibitem{merlin78} R. Merlin, G. G\"untherodt, R. Humphreys, M. Cardona, R. Suryanarayanan,
and F. Holtzberg, Phys. Rev. {\bf B17}, 4951 (1978).

\bibitem{vitins} J. Vitins, J. Magn. Mater. {\bf 5}, 212 (1977) and J. Vitins and P. Wachter,
Physica {\bf B86-88}, 213 (1977).

\bibitem{isobe98} M. Isobe, Y. Ueda, J. Magn. and Magn. Materials {\bf 177-181}, 671 (1998).

\bibitem{ohama98} T. Ohama, H. Yasuoka, M. Isobe, and Y. Ueda,
Phys. Rev. {\bf B 59}, 3299 (1999).

\bibitem{koeppen} M. K\"oppen, D. Pankert, R. Hauptmann, M. Lang, M. Weiden,
C. Geibel, F. Steglich, Phys. Rev. {\bf B57}, 8466
(1998).

\bibitem{fertey} F. Fertey, M. Poirier, M. Castonguay, J. Jegoudez, A. Revcolevschi, Phys.
Rev. {\bf B57}, 13698 (1998).

\bibitem{sherman} This qualitative approach gives a big ratio $\Delta Q/\overline{Q}$.
Typically, the phonon-induced charge transfer can be estimated as $\Delta Q/%
\overline{Q}\sim Cz_{0}/W,$ with $C$ is the electron
phonon-coupling constant and $W\sim 1$ eV  is the electron band
width, respectively. Since a
first-principle estimate gives $z_{0}\sim a\left( m/M\right) ^{1/4}$ and $%
C\sim W/a$ ($a$ is the lattice constant and $m$ is the electron mass), $%
\Delta Q/\overline{Q}\sim \left( m/M\right) ^{1/4}\sim 0.1.$
However, in our case $W=2t_{\bot }$, that is rather small, while
for the coupling we assume $C=eE_{\text{a}}$. The electric field
of the oxygen ions surrounding V in the basal palne, diminishes
the coupling. Note that not only the Coulomb forces but also the
overlap of the oxygen and vanadium orbitals contribute to the
value of $C.$ Quantitatively, $C$ can be determined only as a
result of numerical bandstructure calculations. We suppose,
however, that $C$ is large enough since the asymmetry of the
elementary cell is large.


\bibitem{sherman2} E.Ya. Sherman, Solid State Commun. {\bf 104},
619 (1997).

\bibitem{sherman3}E. Ya. Sherman, et al., to be published.

\bibitem{wochner} P. Wochner, J.M. Tranquada, D.J. Buttrey, and V.
Sachan, Phys. Rev. {\bf B 57}, 1066 (1998).

\bibitem{lee97} S.-H. Lee and S.-W. Cheong, Phys. Rev. Lett {\bf 79}, 2514 (1997).



%
%
%
%
%
%
%
%
%
%
%
%
%
%



\end{thebibliography}
\end{document}